\documentclass[12pt]{article}
\usepackage{graphicx}
\usepackage{url}

\begin{document}

\section*{Transferability of crystal-field parameters for rare-earth
  ions in Y$_2$SiO$_5$ tested by Zeeman spectroscopy}

\begin{center}
N. L. Jobbitt$^{a,b}$,
S. J. Patchett$^{a,b}$, 
Y. Alizadeh$^{a,b}$,
M. F. Reid$^{a,b,*}$, \\
J.-P. R.  Wells$^{a,b}$,
S. P. Horvath$^{a,c}$,
J. J. Longdell$^{b,c}$, \\
A. Ferrier$^{d,e}$
and
P. Goldner$^d$

\bigskip
\noindent
$^{a}$School of Physical and Chemical Sciences, University of Canterbury, \\
PB 4800, Christchurch 8140, New Zealand
\\
$^{b}$The Dodd-Walls Centre for Photonic and Quantum Technologies,\\
New Zealand
\\
$^{c}$Department of Physics, University of Otago, PB 56, Dunedin,\\
 New Zealand
\\
$^{d}$Chimie ParisTech, PSL University, CNRS, \\ Institut de Recherche de Chimie
Paris, Paris, France\\
$^{e}$Facult\'{e} des Sciences et Ing\'{e}nierie, Sorbonne Universit\'{e}, 
Paris, France

\smallskip

$^*$Email: \url{mike.reid@canterbury.ac.nz}

\bigskip

\noindent
\today
\end{center}


\bigskip\noindent \textbf{Abstract---} Zeeman spectroscopy is used to
demonstrate that phenomenological crystal-field parameters determined
for the two $C_1$ point-group sites in Er$^{3+}$:Y$_2$SiO$_5$ may be
transferred to other ions.  The two crystallographic six- and
seven-coordinate substitutional sites may be distinguished by comparing
the spectra with crystal-field calculations.

\bigskip

\newpage

\section{Introduction}

Yttrium orthosilicate (Y$_2$SiO$_5$) doped with rare-earth ions
has been widely used over the past decade in the development of
quantum-information devices.  Yttrium has
a very small nuclear magnetic moment, while isotopes of Si and O with
non-zero nuclear spin have very low natural abundances. This minimises
decoherence due to spin flips, giving outstanding coherence
properties. Furthermore, the rare-earth substitutional sites in
Y$_2$SiO$_5$ has $C_1$ point-group symmetry, giving highly admixed
wavefunctions and enabling efficient and diverse optical pumping schemes
\cite{rippe2005,lauritzen2008,longdell_experimental_2004}.  

Performance improvements for these devices rely on accurate modelling of
magnetic-hyperfine structure. For example, the spin Hamiltonian for
Eu$^{3+}$:Y$_2$SiO$_5$ was utilized in a computational search for
magnetic field orientations exhibiting a near-zero gradient with respect
to hyperfine energy levels. This Zero-First-Order-Zeeman (ZEFOZ)
technique enabled an experimental demonstration of a coherence time of
six hours in $^{151}$Eu$^{3+}$:Y$_2$SiO$_5$ \cite{zhong_optically_2015}.

Spin Hamiltonians are not transferable between electronic levels of the
same ion, or to other ions. This limits the opportunity to explore
candidate systems by theoretical modelling. On the other hand,
crystal-field calculations
\cite{carnall_systematic_1989,GoBi96,NeNg00,liu_electronic_2006} model
the electronic structure of the entire $4f^N$ configuration. The
parameters show systematic trends across the rare-earth series, so they
may be transferred from ion to ion. Crystal-field calculations may be
used to construct spin Hamiltonians, or used directly for
magnetic-hyperfine calculations.  Since the crystal-field Hamiltonian
automatically handles mixing of crystal-field levels by a magnetic
field, crystal-field calculations may be used to extend the ZEFOZ method
to large magnetic fields where a simple spin Hamiltonian approach breaks
down. This is particularly relevant given the demonstration of a
coherence time exceeding one second in Er$^{3+}$:Y$_2$SiO$_5$ using a 7
Tesla magnetic field by Ran\u{c}i\'{c} et al.\ \cite{rancic2018}.

Determination of crystal-field parameters for low-symmetry systems is
non-trivial. Data from Zeeman splitting is essential to give orientation
information necessary to determine a unique set of parameters
\cite{antipin1972,smcaf22018}.  In $C_1$ symmetry (i.e.\ no symmetry)
there are 27 crystal-field parameters, which makes calculations
computationally challenging.  Previous work on $C_1$ symmetry sites has
been based on \emph{ab initio} calculations
\cite{Avanesov1992,doualan_energy_1995,wen2014}, or the use of a
higher-symmetry approximation to reduce the number of parameters
\cite{guillot-noel_calculation_2010,sukhanov2018}.

We have recently developed techniques that make full phenomenological
crystal-field fits for $C_1$ point-group symmetry sites practical. These
methods have been applied to both sites of Er$^{3+}$:Y$_2$SiO$_5$
\cite{Horvath2016,erysocf2018}. The fits used both optical,
magneto-optical and electron-paramagnetic resonance experimental data
from the literature
\cite{doualan_energy_1995,sun_magnetic_2008,chen2018}.

In this work we demonstrate the transferability of the crystal-field
parameters to the ions Sm$^{3+}$ and Nd$^{3+}$ by performing Zeeman
spectroscopy. Calculations based on the Er$^{3+}$ parameters clearly
distinguish the two sites, and this information may be combined with EPR
and ab-initio calculations for particular ions to allow the
identification of sites.

\section{Experimental and theoretical techniques}

Y$_2$SiO$_5$ (in the X2 phase) is a monoclinic crystal with $C^6_{2h}$
space group symmetry. The yttrium ions occupy two crystallographically
distinct sites, each with $C_1$ point-group symmetry, referred to as
site 1 and site 2, corresponding to oxygen coordination numbers of six
and seven, respectively \cite{maksimov1971crystal}. Y$_2$SiO$_5$ has
three perpendicular optical-extinction axes: the crystallographic $b$
axis, and two mutually perpendicular axes labelled $D_1$ and $D_2$. In
our calculations we follow the convention of identifying these as the
$z$, $x$, and $y$ axes respectively \cite{sun_magnetic_2008}.

Samples of Y$_2$SiO$_5$ doped with Er$^{3+}$ (50\,ppm) and Nd$^{3+}$ (200\,ppm)
were prepared in Paris. The Sm$^{3+}$ (5000\,ppm) sample was supplied by
Scientific Materials.  All samples were oriented using Laue
backscattering. The samples were cuboids with the $D_1$ and $D_2$ and
$b$ axes through the faces and dimensions of approximately 5\,mm. Infrared
spectroscopy was performed using a 0.075\,cm$^{-1}$ resolution Bruker
Vertex 80 with an optical path purged by N$_2$ gas.  Zeeman spectroscopy
was performed using a 4\,T, simple solenoid, superconducting magnet with
samples cooled by thermal contact with a copper sample holder fixed
through the centre of the solenoid.  Measurements were carried out at
4.2\,K.

The Hamiltonian appropriate for
modelling the $4f^N$ configuration is
\cite{carnall_systematic_1989,GoBi96,liu_electronic_2006}
\begin{equation}
  H = H_{\mathrm{FI}} + H_{\mathrm{CF}} + H_{\mathrm{Z}} + H_{\mathrm{HF}}.
  \label{eqn:hdefn}
\end{equation} 
The terms in this equation represent
the free-ion contribution, the crystal-field interaction, the Zeeman
term, and the  electron-nuclear hyperfine interaction.

The free-ion Hamiltonian may be written as
\begin{eqnarray}
  H_{\mathrm{FI}} &=& E_\mathrm{avg} + \sum_{k=2,4,6} F^k f_k + \zeta A_{\mathrm{SO}} + \alpha L(L+1) + \gamma G(R_7) \nonumber \\ 
  &+& \beta G(G_2) + \sum_{i = 2,3,4,6,7,8} T^i t_i + \sum_{i=0,2,4} M^i m_i + \sum_{i=2,4,6} P^i p_i.
  \label{eqn:hfidefn}
\end{eqnarray}
$E_\mathrm{avg}$ is a constant configurational shift, $F^k$ Slater
parameters characterizing aspherical electrostatic repulsion, and
$\zeta$ the spin-orbit coupling constant. The other terms 
parametrize two- and three- body interactions, as well as higher-order
spin-dependent effects
\cite{carnall_systematic_1989,liu_electronic_2006}. 

The crystal-field Hamiltonian has the form
\begin{equation}
  H_{\mathrm{CF}} = \sum_{k,q} B^k_q C^{(k)}_q,
  \label{eqn:hcfc1}
\end{equation}
for $k = 2, 4, 6$ and $q = -k \cdots k$. The $B^k_q$ parameters are the
crystal-field expansion coefficients and C$^{(k)}_q$ are spherical
tensor operators. In $C_1$ symmetry all non-axial ($q \neq
0$) $B^k_q$ parameters are complex, leading to a total of 27 parameters. 

We do not explicitly consider hyperfine interactions in this work. The
reader is referred to Refs.\ \cite{Horvath2016,smcaf22018,erysocf2018}
for a discussion of the calculation of hyperfine effects with a crystal-field
model.

\begin{figure}[tb!]
\begin{center}
\includegraphics[width=0.5\linewidth]{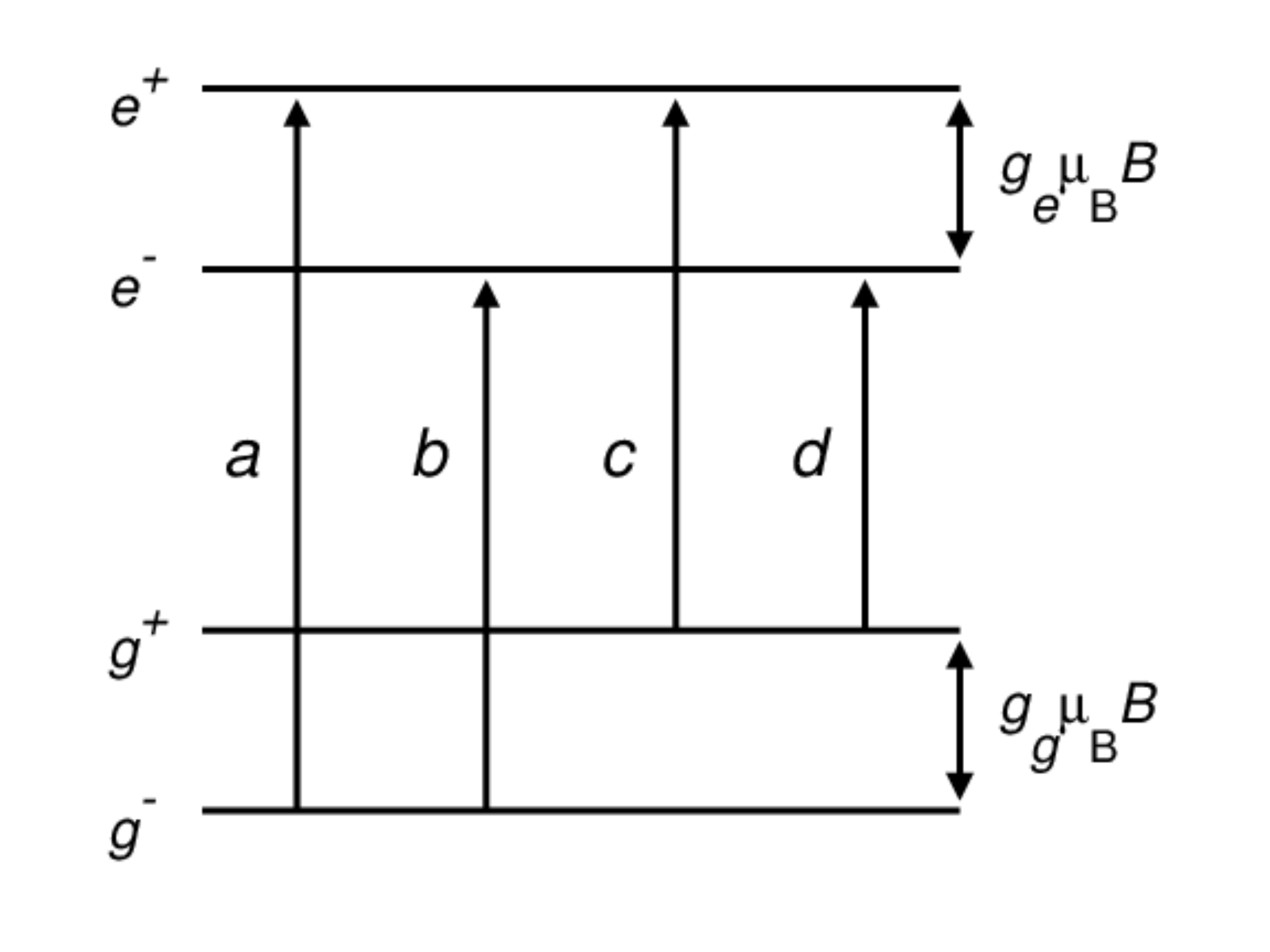}
\end{center}
\caption{\label{fig:zeeman} Schematic diagram of Zeeman splitting of
  ground ($g^\pm$) and excited ($e^\pm$) electronic states of a Kramers
  ion, and the possible transitions.  $g_g$ and $g_e$ are the $g$ values
  of the ground and excited states.  }
\end{figure}

The Zeeman interaction for an external magnetic field is represented by
\begin{equation}
  H_{\mathrm{Z}} = \mu_{\mathrm{B}}  \mathbf{B} \cdot (\mathbf{L} + 2 \mathbf{S}),
  \label{eqn:hZdefn}
\end{equation}
where $\mu_{\mathrm{B}}$ is the Bohr magneton, and $\mathbf{L}$ and
$\mathbf{S}$ are the total orbital and spin angular momenta. In this
work we only consider Kramers ions (with total spin a multiple of
$\hbar/2$). For Kramers ions in low-symmetry sites all electronic levels
are doubly-degenerate in the absence of an applied magnetic field. For
low magnetic fields the splitting of these doublets may be parametrized
by $g$ values, such that the splitting $\Delta_E$ is a function of the
magnetic field $\mathbf{B}$ applied in a particular direction:
\begin{equation}\label{eq:gvalues}
\Delta_E = g \mu_{\mathrm{B}} |\mathbf{B}|. 
\end{equation}
Transitions between the ground electronic state and an excited
electronic state will, in general, contain four spectral lines, as
indicated schematically in Figure \ref{fig:zeeman}, with the energy
differences depending on sums and differences of ground and
excited-state $g$ values.  Measurements of these differences can reveal
a large amount of information regarding the wavefunctions and the effect
of magnetic fields using moderately high-resolution spectroscopy, such
as Fourier-transform absorption spectroscopy. In contrast, non-Kramers
ions in low-symmetry sites have no electronic degeneracy and
measurements of magnetic splitting of the hyperfine structure usually
requires high-resolution techniques, such as in Ref.\
\cite{zhong_optically_2015}.

Parameters for Site 1 and Site 2 of Er$^{3+}$:Y$_2$SiO$_5$ (using the
convention of Ref.\ \cite{sun_magnetic_2008}) are given by Horvath
\cite{Horvath2016}. An alternative parameter set for Site 1 is given in
Ref.\ \cite{erysocf2018}. However, since the latter is optimised for
high-resolution magneto-hyperfine data, we use the former in this work,
as this gives more consistency between the parameter sets for the two
sites. Since the ionic radius and the crystal-field parameters reduce
across the rare-earth series
\cite{carnall_systematic_1989,GoBi96,NeNg00,liu_electronic_2006}, the
crystal-field parameters for Sm$^{3+}$ and Nd$^{3+}$ should be scaled up
from the Er$^{3+}$ parameters. However we found that for the levels
considered here, scaling made a very small difference, and the
calculations presented here use unscaled parameters.  The free-ion
parameters for Sm$^{3+}$ and Nd$^{3+}$ were taken from Ref.\
\cite{carnall_systematic_1989}.

\section{Results and discussion}

Zeeman spectra for Nd$^{3+}$, Sm$^{3+}$, and Er$^{3+}$ in Y$_2$SiO$_5$
crystals are given in Figure \ref{fig:ysozeeman}. Calculated magnetic
splittings for the particular orientation and field chosen are indicated
on the diagrams. Details of the transitions and magnetic fields are
given in Table \ref{tab:ysozeeman}, along with calculated $g$ values.

\subsection{ Er$^{3+}$:Y$_2$SiO$_5$}

We begin with Er$^{3+}$:Y$_2$SiO$_5$, since this is the ion for which
the crystal-field parameters were determined
\cite{Horvath2016,erysocf2018}.  Er$^{3+}$ has 11 $4f$ electrons and has
a reltively small ionic radius, similar to Y$^{3+}$, so it would be
expected to substitute equally into both Y$^{3+}$ sites.

Figure \ref{fig:ysozeeman}(c) shows the spectrum of
Er$^{3+}$:Y$_2$SiO$_5$ in the region of the transitions from the ground
state to the lowest-energy $^4$I$_{13/2}$ states for each site, with a 1\,T field along
the $b$ axis. The magnetic splittings of these transitions were
extensively studied by Sun et al.\ \cite{sun_magnetic_2008} and our
spectrum corresponds to the the $\theta=0$ points of their Figures 3(c)
and 4(c). The data from that study was crucial input to the
crystal-field fit \cite{Horvath2016,erysocf2018}. The magnetic
splittings of other transitions are also generally in good agreement,
confirming the experimental energy-level assignments used in the fits. 

The absorption for the Site 1 and Site 2 transitions is comparable, so
if the oscillator strengths are assumed to be similar then the
concentration of Er$^{3+}$ in each site is comparable. The crystal-field
splitting for Site 1 is larger than for Site 2
\cite{doualan_energy_1995}, and this is reflected in our fitted
crystal-field parameters \cite{Horvath2016}. This suggests that Site 1
is the six-fold coordinate site, for which the crystal-field parameters
are calculated to be larger \cite{wen2014}.

\subsection{ Nd$^{3+}$:Y$_2$SiO$_5$}

Nd$^{3+}$ has 3 $4f$ electrons, so has a larger ionic radius than
Er$^{3+}$ and substitution into the seven-coordinate site is expected to
be favoured. The site with the largest absorbance, labelled Type 1 in
Ref. \cite{beach1990}, has been studied by EPR
\cite{wolfowicz2015,sukhanov2018}.  In Ref.\
\cite{sukhanov2018} this site was identified as
seven-coordinate by pulsed EPR measurements.

Figure \ref{fig:ysozeeman}(a) shows the spectrum of
Nd$^{3+}$:Y$_2$SiO$_5$ in the region of the transitions from the ground state to the
lowest-energy $^4$F$_{3/2}$ states for each site, with a 2\,T field along the $b$
axis. The magnetic splitting for the site with the strongest
absorption matches the Site 2 calculation. The spectrum, and the calculated ground-state $g$ value of
4 for a field along the $b$ axis is consistent with the EPR data.

\subsection{ Sm$^{3+}$:Y$_2$SiO$_5$}

Sm$^{3+}$ is a Kramers ion with a large number of absorption lines in
the IR, visible, and UV spectral regions. It is, therefore, an ideal ion
for extensive Zeeman measurements. 

\begin{figure}[tb!]
\begin{center}
\vspace{-2cm}
\includegraphics[width=0.55\linewidth]{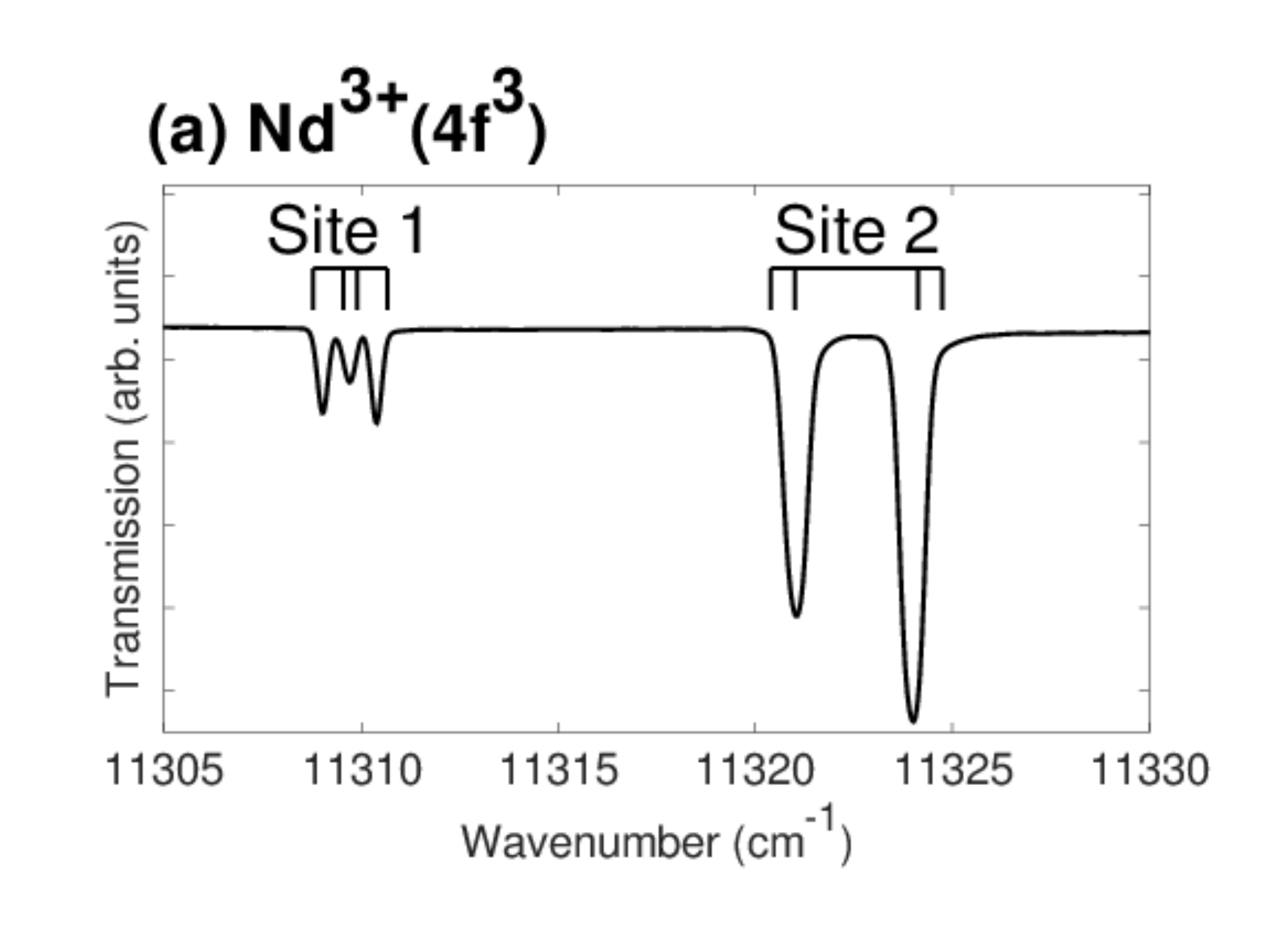}\\
\vspace{-0.5cm}
\includegraphics[width=0.55\linewidth]{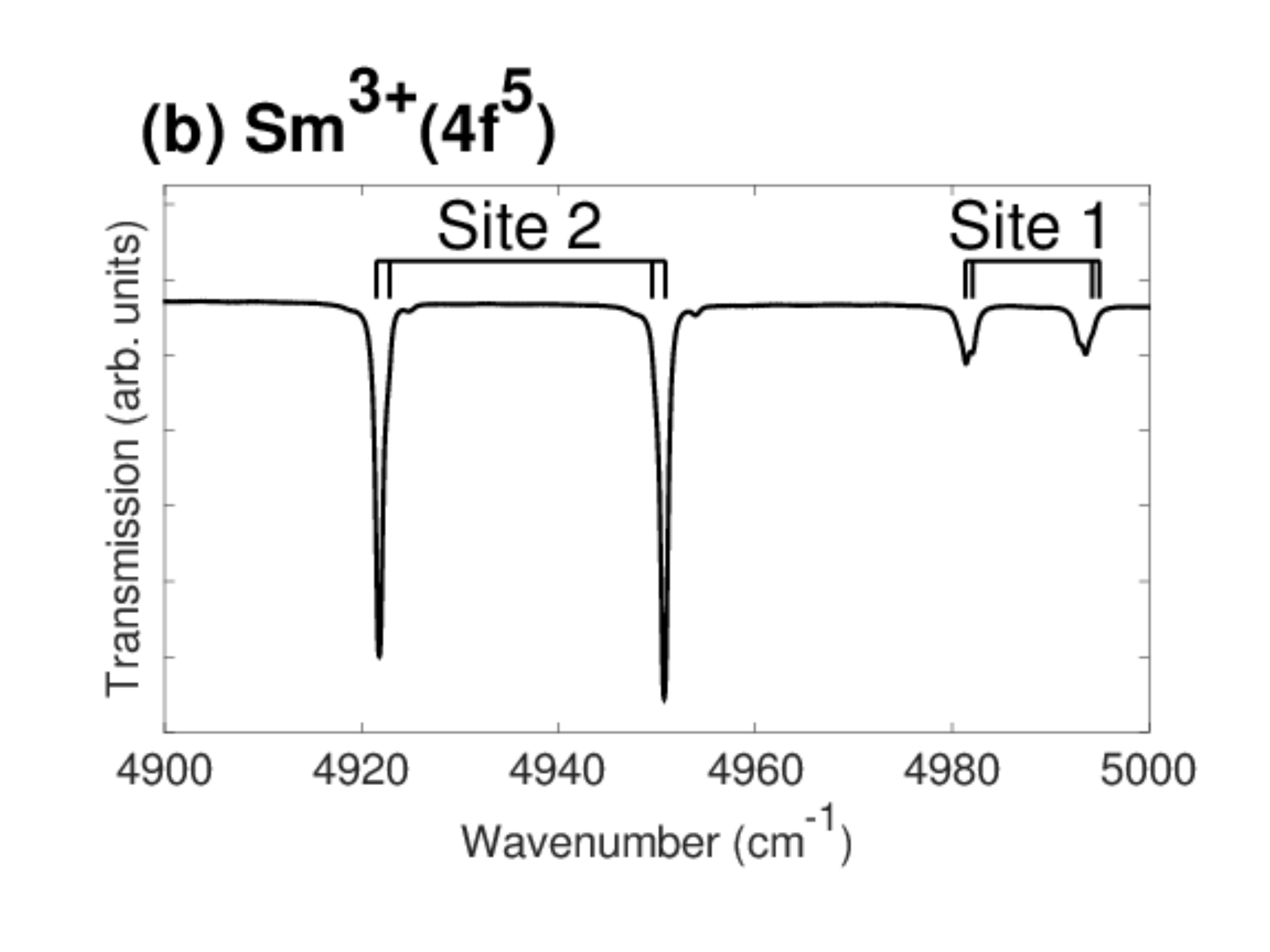}\\
\vspace{-0.5cm}
\includegraphics[width=0.55\linewidth]{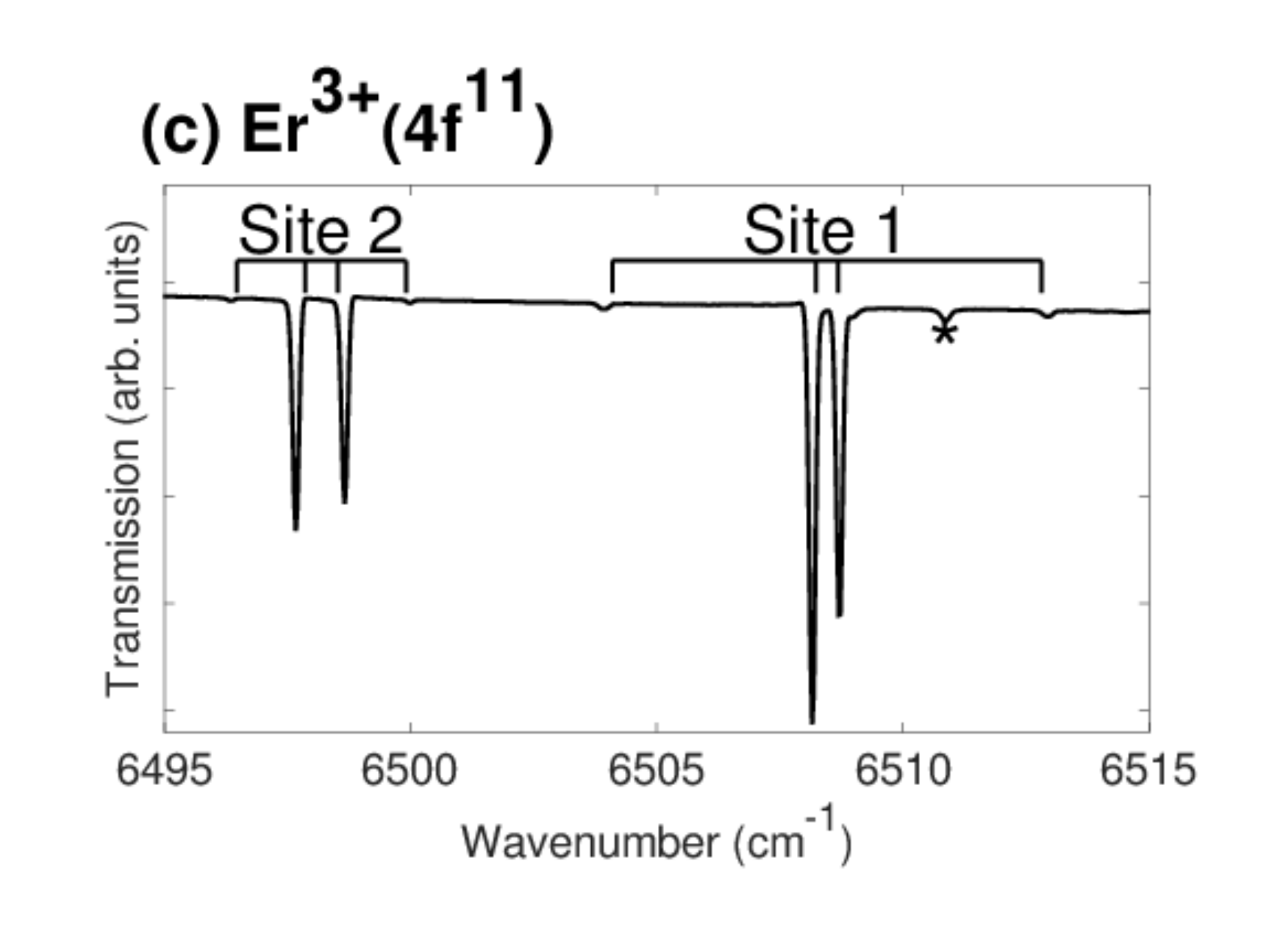}\\
\vspace{-0.5cm}

\caption{\label{fig:ysozeeman} Zeeman spectra for Nd$^{3+}$, Sm$^{3+}$,
  and Er$^{3+}$ in Y$_2$SiO$_5$. Transitions from the ground state to
  the lowest-energy states of a particular excited multiplet for each
  site are shown. Details of the excited multiplets, and the magnetic field
  directions and strengths, are given in Table \ref{tab:ysozeeman}.
  Zeeman splittings calculated from the crystal-field model are shown in
  each plot. All measurements were done at 4.2\,K.  For
  Sm$^{3+}$:Y$_2$SiO$_5$, the small features on the high-energy side of
  the site 2 lines are satellite lines.  For Er$^{3+}$:Y$_2$SiO$_5$, the
  feature marked with an asterisk is not associated with Er$^{3+}$.}
\end{center}
\end{figure}

Sm$^{3+}$ has 5 $4f$ electrons, so it has an ionic radius slightly
smaller than Nd$^{3+}$ and, again, substitution into the seven-coordinate
site is expected to be favoured.  Figure \ref{fig:ysozeeman}(b) shows
the spectrum of Sm$^{3+}$:Y$_2$SiO$_5$ in the region of the transitions from the
ground state to the lowest-energy $^6$H$_{13/2}$ states for each site, with a 4\,T
field along the $D_1$ axis.

The magnetic splitting for the transitions with highest absorbance is
consistent with the Site 2 calculation. The $g$ values for the ground
state for both sites are small (Table \ref{tab:ysozeeman}) and are not
fully resolved in Figure \ref{fig:ysozeeman}(b), whereas the
excited-state $g$ values are large, particularly for Site 2.




\section{Conclusions}

We have used Zeeman spectroscopy to demonstrate that crystal-field
parameters determined for Er$^{3+}$ in the two Y$_2$SiO$_5$
substitutional sites give a reasonable account of magnetic splittings
in other ions. Future work will use more extensive measurements to
refine the crystal-field parameters across the rare-earth series. This
will provide improved modelling relevant to the development of
quantum-information applications.

\clearpage

\begin{table}[tb!]
  \caption{\label{tab:ysozeeman}
    Details of the transitions and magnetic-field settings for the spectra of 
    Nd$^{3+}$, Sm$^{3+}$, and Er$^{3+}$ in
    Y$_2$SiO$_5$ given in Figure \ref{fig:ysozeeman}. 
    For each site the experimental wavenumber at zero magnetic field ($E_0$) is listed, and 
    $g$ values
    calculated from the crystal-field model, using the notation
    of Figure \ref{fig:zeeman}. 
  }
\begin{center}
\begin{tabular}{cccccccccc}
    \hline \hline
    Ion &Excited  &\multicolumn{2}{c}{Magnetic Field} &\multicolumn{3}{c}{Site 1} &\multicolumn{3}{c}{Site 2}\\
        &multiplet &axis &$B$ (T)   &$E_0$  (cm$^{-1}$)   &$g_g$    &$g_e$ &$E_0$ (cm$^{-1}$)   &$g_g$   &$g_e$\\
    \hline
Nd$^{3+}$ & $^4$F$_{3/2}$  &$b$   &2.0 &11309.7   &1.20     &0.82  &11322.5    &4.00   &0.66\\

Sm$^{3+}$ & $^6$H$_{13/2}$ &$D_1$ &4.0 &4987.8    &0.39      &6.36  &4936.3    &0.71   &15.00\\ 

Er$^{3+}$ & $^4$I$_{13/2}$ &$b$   &1.0  &6508.4  &7.82      &10.00 &6498.1    &3.04   &4.39\\
   \hline
 \end{tabular}
\end{center}
\end{table}

\end{document}